\documentclass{aastex62}
\usepackage{amsmath,amssymb,CJK}

\usepackage{soul}
\received{--}
\revised{--}
\accepted{--}
\submitjournal{ApJL}

\shorttitle{Active Asteroid (6478) Gault}
\shortauthors{Ye et al.}

\begin{document}
\begin{CJK*}{UTF8}{gbsn}

\title{Multiple Outbursts of Asteroid (6478) Gault\footnote{Data and codes that generate the figures of this work are available on\dataset[10.5281/zenodo.2590809]{10.5281/zenodo.2590809} and \url{https://github.com/Yeqzids/activation_of_6478_gault}.}
}

\correspondingauthor{Quanzhi Ye}
\email{qye@caltech.edu}

\author[0000-0002-4838-7676]{Quanzhi Ye (叶泉志)}
\affiliation{Division of Physics, Mathematics and Astronomy, California Institute of Technology, Pasadena, CA 91125, U.S.A.}
\affiliation{Infrared Processing and Analysis Center, California Institute of Technology, Pasadena, CA 91125, U.S.A.}

\author[0000-0002-6702-7676]{Michael S. P. Kelley}
\affiliation{Department of Astronomy, University of Maryland, College Park, MD 20742, U.S.A.}

\author[0000-0002-2668-7248]{Dennis Bodewits} 
\affiliation{Department of Physics, Auburn University, Auburn, AL 36849, U.S.A.}

\author[0000-0002-4950-6323]{Bryce Bolin}
\altaffiliation{B612 Asteroid Institute and DIRAC Institute Postdoctoral Fellow}
\affiliation{DIRAC Institute, Department of Astronomy, University of Washington, 3910 15th Avenue NE, Seattle, WA 98195, U.S.A.}
\affiliation{B612 Asteroid Institute, 20 Sunnyside Ave, Suite 427, Mill Valley, CA 94941, U.S.A.}

\author{Lynne Jones}
\affiliation{DIRAC Institute, Department of Astronomy, University of Washington, 3910 15th Avenue NE, Seattle, WA 98195, U.S.A.}

\author{Zhong-Yi Lin \begin{CJK*}{UTF8}{bkai}(林忠義)\end{CJK*}}
\affiliation{Institute of Astronomy, National Central University, 32001, Taiwan}

\author[0000-0001-8018-5348]{Eric C. Bellm}
\affiliation{DIRAC Institute, Department of Astronomy, University of Washington, 3910 15th Avenue NE, Seattle, WA 98195, U.S.A.}

\author[0000-0001-5060-8733]{Richard Dekany}
\affiliation{Caltech Optical Observatories, California Institute of Technology, Pasadena, CA 91125, U.S.A.}

\author[0000-0001-5060-8733]{Dmitry A. Duev}
\affiliation{Division of Physics, Mathematics and Astronomy, California Institute of Technology, Pasadena, CA 91125, U.S.A.}

\author{Steven Groom}
\affiliation{Infrared Processing and Analysis Center, California Institute of Technology, Pasadena, CA 91125, U.S.A.}

\author{George Helou}
\affiliation{Infrared Processing and Analysis Center, California Institute of Technology, Pasadena, CA 91125, U.S.A.}

\author[0000-0001-5390-8563]{Shrinivas R. Kulkarni}
\affiliation{Division of Physics, Mathematics and Astronomy, California Institute of Technology, Pasadena, CA 91125, U.S.A.}

\author[0000-0002-6540-1484]{Thomas Kupfer}
\affiliation{Kavli Institute for Theoretical Physics, University of California, Santa Barbara, CA 93106, U.S.A.}

\author{Frank J. Masci}
\affiliation{Infrared Processing and Analysis Center, California Institute of Technology, Pasadena, CA 91125, U.S.A.}

\author{Thomas A. Prince}
\affiliation{Division of Physics, Mathematics and Astronomy, California Institute of Technology, Pasadena, CA 91125, U.S.A.}

\author[0000-0001-6753-1488]{Maayane T. Soumagnac}
\affiliation{Department of Particle Physics and Astrophysics, Weizmann Institute of Science 
234 Herzl St., Rehovot, 76100, Israel}



\begin{abstract}

Main-belt asteroid (6478) Gault unexpectedly sprouted two tails in late 2018 and early 2019, identifying it as a new active asteroid. Here we present observations obtained by the 1.2-m Zwicky Transient Facility survey telescope that provide detailed time-series coverage of the onset and evolution of Gault's activity. Gault exhibited two brightening events, with the first one starting on 2018 Oct. 18$\pm5$~days and a second one starting on 2018 Dec. 24$\pm1$~days. The amounts of mass released are $2\times10^7$~kg and $1\times10^6$~kg, respectively. Based on photometric measurements, each event persisted for about a month. Gault's color has not changed appreciably over time, with a pre-outburst color of $g_\mathrm{PS1}-r_\mathrm{PS1}=0.50\pm0.04$ and $g_\mathrm{PS1}-r_\mathrm{PS1}=0.46\pm0.04$ during the two outbursts. Simulations of dust dynamics shows that the ejecta consists of dust grains of up to $10~\micron$ in size that are ejected at low velocities below $1~\mathrm{m~s^{-1}}$ regardless of particle sizes. This is consistent with non-sublimation-driven ejection events. The size distribution of the dust exhibits a broken power-law, with particles at 10--20~$\micron$ following a power-law of $-2.5$ to $-3.0$, while larger particles follow a steeper slope of $-4.0$. The derived properties can be explained by either rotational excitation of the nucleus or a merger of a near-contact binary, with the latter scenario to be statistically more likely.

\end{abstract}



\section{Introduction} \label{sec:intro}

The active asteroids is a small-body population in our Solar System that are dynamically asteroids, but can transiently or periodically display comae and tails. While the activity of comets is driven by sublimation, asteroid activity may be driven by a diverse set of mechanisms, including sublimation, rotational instability, impacts, and thermal fracturing \citep{Jewitt2015}. The occurrence rate of active asteroids is about 1--10 per 100,000 \citep[c.f.][]{Chandler2018}, implying $<100$ active asteroids among the known $\sim$800,000 asteroids.

(6478) Gault is a km-sized asteroid in the Phocaea family \citep{Nesvorny2015} with no previous record of activity. With a Tisserand parameter with respect to Jupiter, $T_\mathrm{J}=3.46$\footnote{From JPL solution \#21, retrieved on 2019 February 10.}, Gault is dynamically a main-belt asteroid. Its activity was first detected by the Asteroid Terrestrial-impact Last Alert System (ATLAS) on 2018 December 8 \citep{Smith2019}, with confirmations by other observers \citep[e.g.][]{Hale2019, Jehin2019, Ye2019}. 

We identified serendipitous observations of Gault in the Zwicky Transient Facility (ZTF) survey data, dating back to 2017 November 14. These observations provide baseline (inactive) photometry as well as extended coverage of the activation of Gault. In this paper, we analyze the time-series data provided by ZTF to characterize the activity of Gault, and to understand the mechanism that drove its activity.

\section{Observations} \label{sec:obs}

ZTF is a synoptic survey combining the 1.2-m Oschin Schmidt telescope with a new 47~deg$^2$ camera at Palomar Observatory. Using 30-second exposures, ZTF can observe an area of 3760 deg$^2$ in an hour, to a typical $5\sigma$ detection limit of $r_\mathrm{PS1}=20.7$. Most survey observations are taken in the ZTF $g$ and $r$ filters. 
The telescope and its survey strategy are described in \citet{Bellm2019} and \citet{Graham2019}.

We use \texttt{ZChecker}\footnote{\url{https://github.com/mkelley/zchecker/}.}, a software suite originally developed to analyze comet images in ZTF data, to extract and process the ZTF images of Gault \citep{Kelley2019}. \texttt{ZChecker} uses the ephemeris provided by JPL HORIZONS \citep{Giorgini2015} and the survey metadata \citep{Masci2019} to identify observations of known small bodies in the ZTF archive. 

The software combines frames from each night into nightly stacks, and applies temporal filtering to the images to highlight morphological changes over time. Based on \texttt{ZChecker}-identified observations, we derived photometry based on point-source-function (PSF) fitting from the ZTF data reduction pipeline \citep{Masci2019}, and extracted the source nearest the ephemeris position of Gault. Photometric zero-points for the absolute calibration are calculated by the ZTF Science Data System by comparing field-stars with the PANSTARRS Data Release 1 catalogue \citep[c.f.][\S~3.5]{Magnier2013, Masci2019}. 

\begin{figure*}
\includegraphics[width=\textwidth]{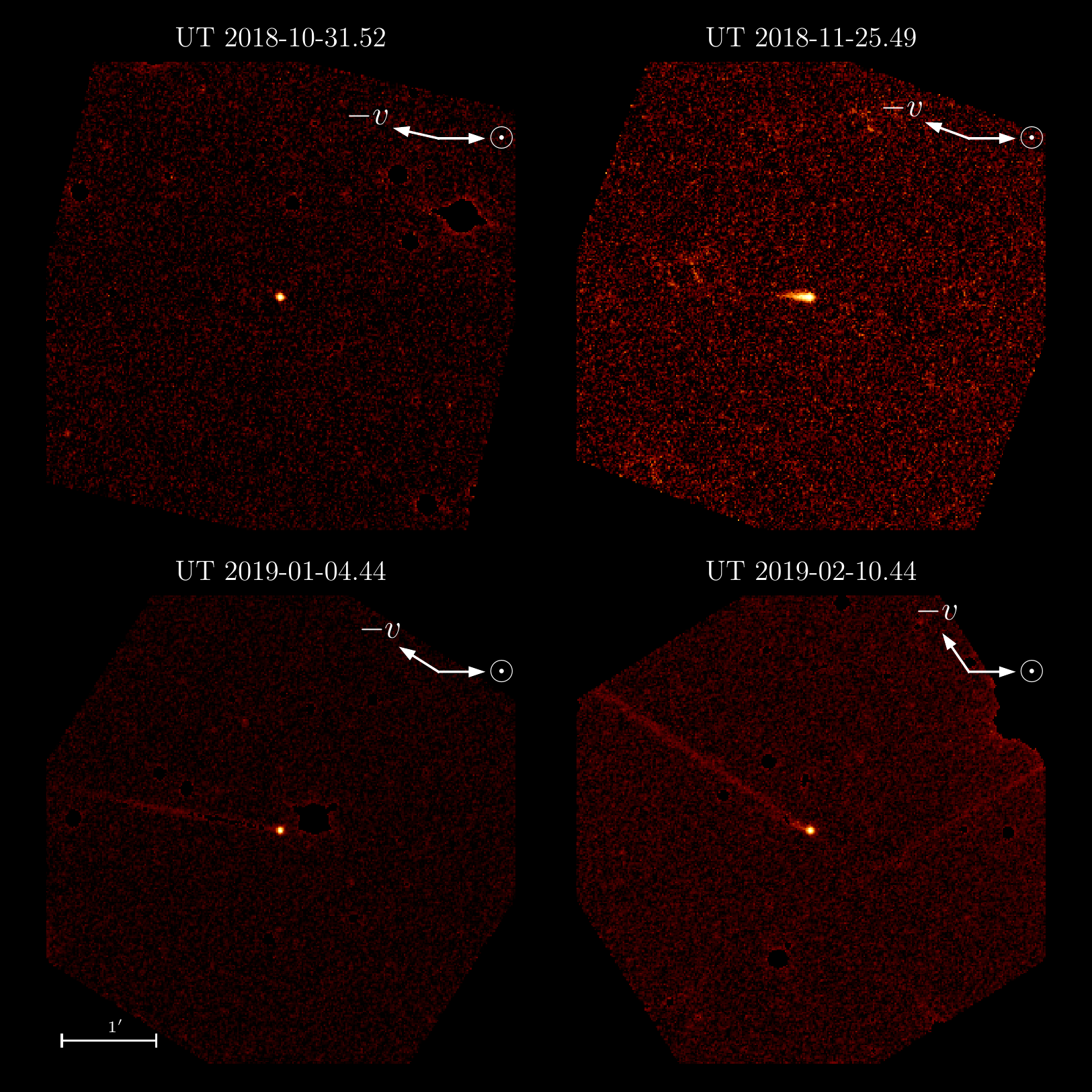}
\caption{ZTF observations of the evolution of Gault's activity from UT 2018 October 31 to 2019 February 10. Most background sources are masked, leaving a few artifacts. Frames are projected into a reference frame with the projected vector to the Sun at $+x$ direction. The $-v$ vector points to the negative of the motion vector. The animation is available in the online Journal. Notebook is available \href{https://github.com/Yeqzids/activation_of_6478_gault/blob/master/img.ipynb}{here}.}  \label{fig:gault}
\end{figure*}

ZTF observed Gault more than 300 times since its commissioning in late 2017, with most observations occurring after October 2018. A time-series animation of ZTF imaging of Gault is shown in Figure~\ref{fig:gault}. The images shown are cut-out frames of $5'\times5'$. Most stars are masked before the frames are median combined into nightly stacks. The total integration time each night varies from 30 seconds to 27 minutes. 

\section{Results} \label{sec:results}

\begin{figure*}
\includegraphics[width=\textwidth]{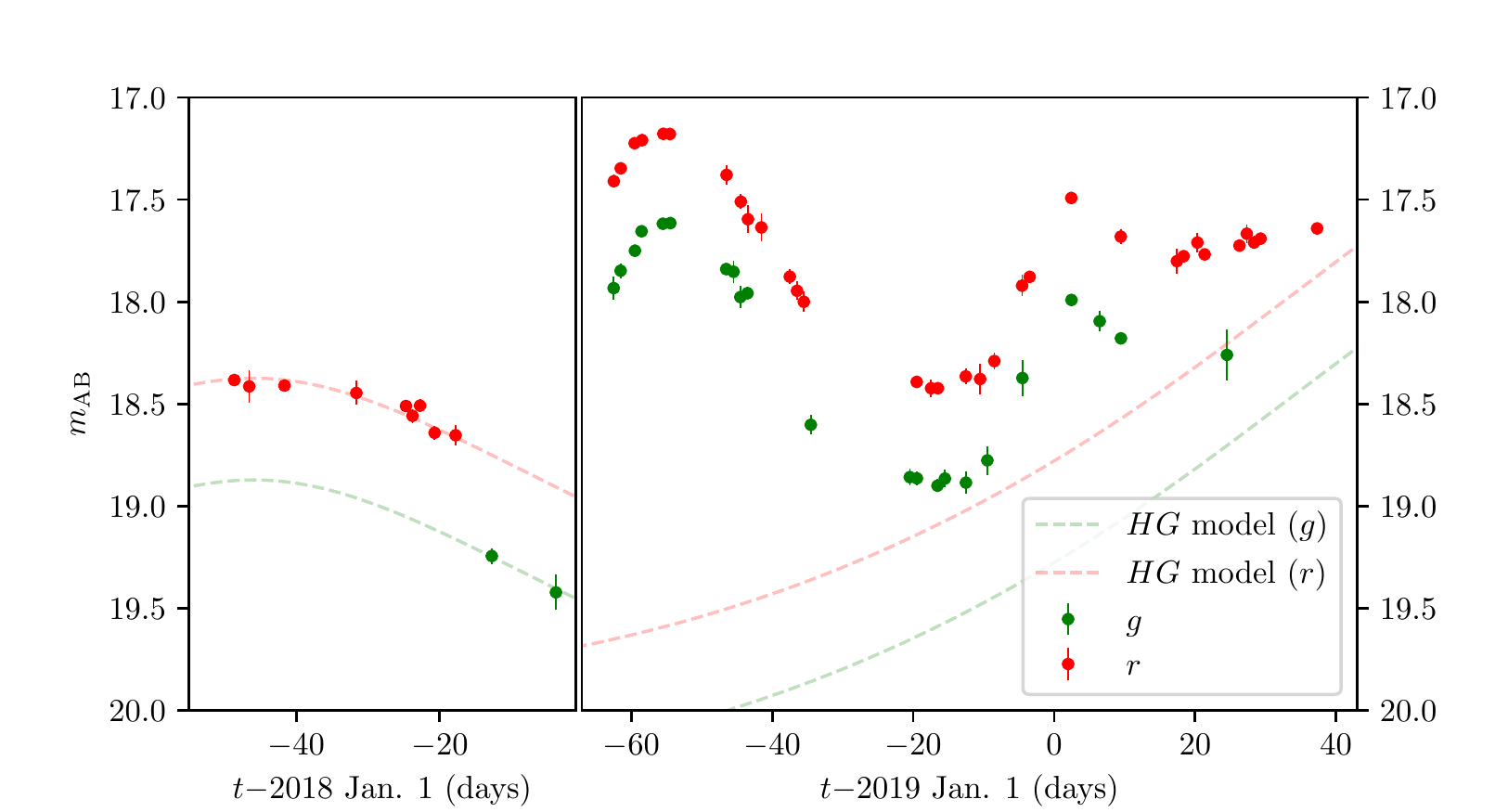}
\caption{Long-term light-curve of Gault in late 2017 (left panel) and late 2018 to early 2019 (right panel) constructed from nightly-binned ZTF photometry. Also shown is the predicted baseline (i.e. inactive) brightness from our $HG$ model. Notebook is available \href{https://github.com/Yeqzids/activation_of_6478_gault/blob/master/lightcurve.ipynb}{here}. \label{fig:lc}}
\end{figure*}

A summary of the ZTF photometry of Gault is shown in Figure~\ref{fig:lc}. We first construct our own $HG$ photometry model for Gault, since absolute magnitudes based on the Minor Planet Center (MPC) database contain biases of up to 0.3--0.5 mag \citep{Juric2002, Pravec2012}. By using the observations acquired in 2017 (pre-outburst), we derive $H_g=14.81\pm0.04$ and $H_r=14.31\pm0.01$, assuming $G=0.15$. This yields $g_\mathrm{PS1}-r_\mathrm{PS1}=0.50\pm0.04$. The color is both broadly consistent with C-type asteroids ($g_\mathrm{PS1}-r_\mathrm{PS1}\approx0.5$) and, to a lesser degree, with S-type asteroids and comets ($g_\mathrm{PS1}-r_\mathrm{PS1}\gtrsim0.6$).

The light-curve shows that Gault was already $\sim2$~mag above baseline when ZTF observed it on UT 2018 October 31.50, the first ZTF observation in 2018. The asteroid was first point-source like in appearance and its size equivalent to the seeing (with full-width-half-maximum, FWHM, of $\sim2''$). The brightness peaked around 2018 November 7 with a final brightness increase of $\Delta$mag$=2.4$. A tail first appeared in ZTF images from UT 2018 November 16. The light curve suggests that the brightening subsided around mid-December 2018.

The photometry shows that a second outburst started between UT 2018 December 24.51 -- 28.46 and peaked near 2019 January 4, with $\Delta$mag$=0.6$ compared to the subsided brightness in mid-December 2018 (accounting for the change of geometry). The brightening lasted about 2 weeks. A new tail was first observed in ZTF images on 2019 January 28. This second tail was observed as early as 2019 January 20 by other observers \citep{Jehin2019, Lee2019}.

During both active phases, Gault had $g_\mathrm{PS1}-r_\mathrm{PS1}=0.46\pm0.04$, in 1$\sigma$ agreement with the color of the nucleus ZTF measured in 2017. The asteroid and its near-nucleus ejecta is thus slightly redder than the Sun \citep[$g_\mathrm{PS1}-r_\mathrm{PS1} = 0.39$; c.f.][]{Willmer2018}. This suggests the dust ejecta does not contain a preponderance of Rayleigh scatterers (i.e. sub-$\micron$-sized dust) or water ice. 




\section{Analysis}

\subsection{Onset of Activity}
\begin{figure*}
\includegraphics[width=\textwidth]{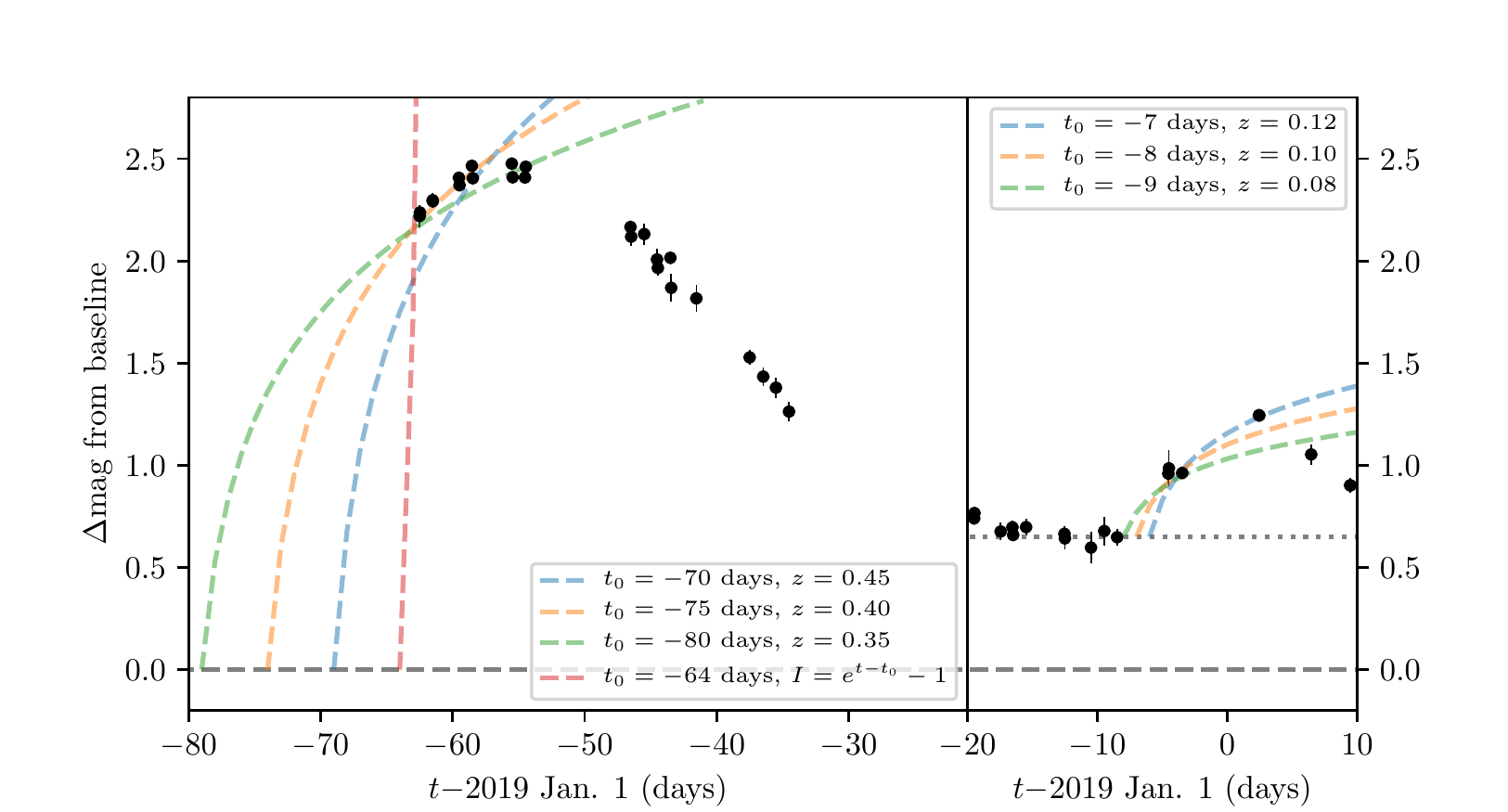}
\caption{Brightness excess of Gault from the prediction by the $HG$ model, with square-law and exponential functions matched to the data points. The left panel shows the first activation, the right panel the onset of the second. The dashed line is the baseline and the dotted line is the asymptote in mid-December 2018. Notebook is available \href{https://github.com/Yeqzids/activation_of_6478_gault/blob/master/lightcurve.ipynb}{here}. \label{fig:lc-norm}}
\end{figure*}

The behavior of brightness over time can be used to determine the onset of activity and to characterize the ejecta. In Figure~\ref{fig:lc-norm} we note that the brightening phase appears to follow a square law:

\begin{equation}
\label{eq:sq-law}
I=z(t-t_0)^2
\end{equation}

\noindent where $I$ is the excess flux, $t-t_0$ is the elapsed time since the start of activity, and $z$ is a constant related to the acceleration of the dust grains. The underlying assumption is that (1) the ejecta is dominated by dust grains that do not fragment; (2) the brightness is proportional to the total surface area of the ejecta; and (3) either the full volume is optically thick until the peak brightness, or the dust is being continually produced for several days. An alternative scenario is that the fragmenting ejecta leads to an exponentially increasing brightness, as seen in cometary outbursts \citep[e.g.][]{Hsieh2010}. The latter scenario cannot describe the light-curve of Gault.

Based on Figure~\ref{fig:lc-norm}, the parameters most compatible with the observations are $t_0 = 2018 $October 18$\pm5$, $z=0.40\pm0.05~\mathrm{day^{-2}}$ for the first outburst, and $t_0=2018$ December 24$\pm1$, $z=0.10\pm0.02~\mathrm{day^{-2}}$ for the second outburst. The spread in $z$ is likely an artifact due to the limited number of observations, especially for the second event.



\subsection{Dust Properties} \label{sec:an:dust}

\subsubsection{General Properties}

To constrain the size of the dust grains in the ejecta and to provide an independent constraint on the onset of activity, we performed a syndyne-synchrone analysis \citep[c.f.][]{Finson1968} of the ZTF images. Syndynes are a set of lines representing the positions of particles of the same sizes, whereas synchrones are a set of lines representing the positions of particles released at the same time. Syndynes are usually expressed using $\beta$, the ratio of the solar radiation and gravitation forces and a proxy for particle size: $\beta=5.74\times10^{-4}/(\rho r)$, where $\rho$ is particle density and $r$ is particle radius, both in MKS units. 

We use images taken on 2018 December 29 and 2019 February 8 for our analysis. These images are selected due to their relatively high signal-to-noise ratios (SNRs). Analysis with a finer time step did not provide extra information, since the event evolves over a timescale of weeks.

\begin{figure*}
\includegraphics[width=\textwidth]{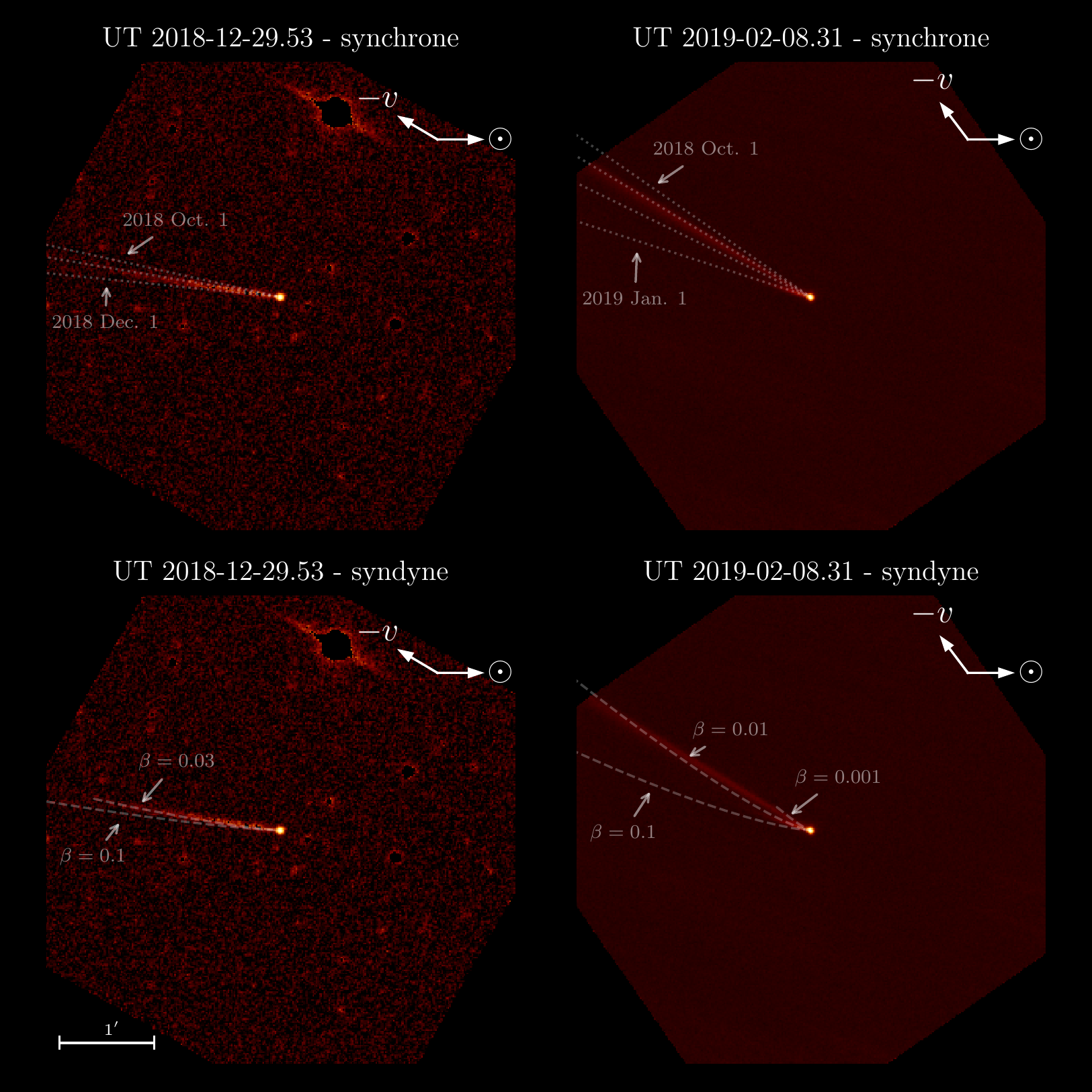}
\caption{ZTF images from 2018 December 29 and 2019 February 8 overlaid with syndyne-synchrone chart. Synchrone lines are the position of particles released on 2018 October 1, November 1, December 1 and 2019 January 1 (for the 2019 February 8 image) in clockwise order. Syndyne lines are calculated for particles released after 2018 October 1. Notebook is available \href{https://github.com/Yeqzids/activation_of_6478_gault/blob/master/syndyne-synchrone.ipynb}{here}. \label{fig:fp}}
\end{figure*}

As shown in Figure~\ref{fig:fp}, the primary tail most closely resembles the synchrone curves that reproduce materials released between late October 2018 and mid November 2018, in agreement with the results above. The new, secondary tail is consistent with an onset around 2019 January 1, also consistent with the second onset date derived above ($\pm1$ week from 2019 January 4).

The range of $\beta$ values (i.e. grain sizes) can be estimated by matching the overlapping sections of the tail and the syndynes. We establish that the smallest grains released in both events have $\beta\approx0.03$ (or approximately 15~$\micron$ in diameter) and the largest grains have at least $\beta=0.001$ (or 0.4~mm in diameter).

The syndyne-synchrone analysis assumes zero terminal ejection speeds of dust grains, and therefore cannot be used to derive ejection speeds of the dust. However, the absence of sun-ward ejecta can be used to constrain the ejection speed \citep[e.g.][]{Hui2017}:

\begin{equation}
\label{eq:v_ej}
v_\mathrm{ej} \leq \frac{(2 \beta G M_\odot \varDelta l)^{1/2}}{r_\mathrm{h}}
\end{equation}

\noindent where $G=6.67\times10^{-11}~\mathrm{N~kg^2~m^2}$ is the gravitational constant, $M_\odot=2\times10^{30}~\mathrm{kg}$ is the mass of the Sun, $\varDelta$ and $r_\mathrm{h}$ are the geocentric and heliocentric distance, and $l\approx1''$ is the apparent sun-ward turnaround distance measured from the nucleus, as constrained by the PSF of our ZTF images. By substituting the variables with corresponding numbers and taking $\beta \leq 0.03$ (from the syndyne-synchrone analysis above), $\varDelta=1.51$~au and $r_\mathrm{h}=2.41$~au for UT 2019 February 8.3, we obtain an upper limit of $v_\mathrm{ej}\leq8~\mathrm{m~s^{-1}}$.

\subsubsection{Dust Tail Model} \label{sec:dust-model} 

The tail morphology can be used to further constrain the properties of the ejecta, and we employ the dust dynamics code originally developed by \citet{Ye2016a}. We focus on the stacked image of 2019 February 8, as it offers the highest SNR in our dataset, with both tails clearly visible.
Although the model contains a number of poorly-constrained parameters and often cannot provide unique solutions, for this particular case, the light-curve and syndyne-synchrone analysis above has provided a set of useful boundary conditions, which helps the identification of improbable solutions.

The model assumes isotropic ejection, with the terminal ejection speed defined as:

\begin{equation}
v_\mathrm{ej} = V_0 \beta^p
\end{equation}

\noindent where $V_0$ is the mean terminal ejection speed of a $\beta=1$ dust grain, and $p$ is a constant. We will test $V_0$ of 1 and 8 $\mathrm{m s^{-1}}$ for $p=0$, and $V_0=50~\mathrm{m s^{-1}}$ for $p=0.5$. These values are chosen because (1) the range of $V_0$ is determined by the largest $V_0$ capable to launch particles from 30~$\micron$ to 1~mm into terminal speed, while not exceeding the $8~\mathrm{m~s^{-1}}$ limit derived above; and (2) it has been found that non-sublimation-driven active asteroids follow $p=0$, while sublimation-driven active asteroids follow $p=0.5$ \citep{Moreno2012,Jewitt2015}. We assume the dust size follows a simple power-law, with a differential size index of $q$. The range of $q$ to be tested is from $-4.0$ to $-2.5$, a range identified by previous observations of comets and active asteroids \citep[c.f.][]{Fulle2004,Jewitt2015,Rotundi2015}, with an interval of 0.5. The $\beta$ of the test particles ranges from 0.0003 to 0.03 as indicated by Figure~\ref{fig:fp}. Particles are released during 2018 November 7$\pm5$ and 2019 January 4$\pm5$, the full-width-half-maxima estimated from Figure~\ref{fig:lc-norm}. We assume the ratio between the dust productions of the two events is $16:1$, in concordance with the amounts of mass losses derived later in \S~\ref{sec:an:mass}.

\begin{figure*}
\includegraphics[width=\textwidth]{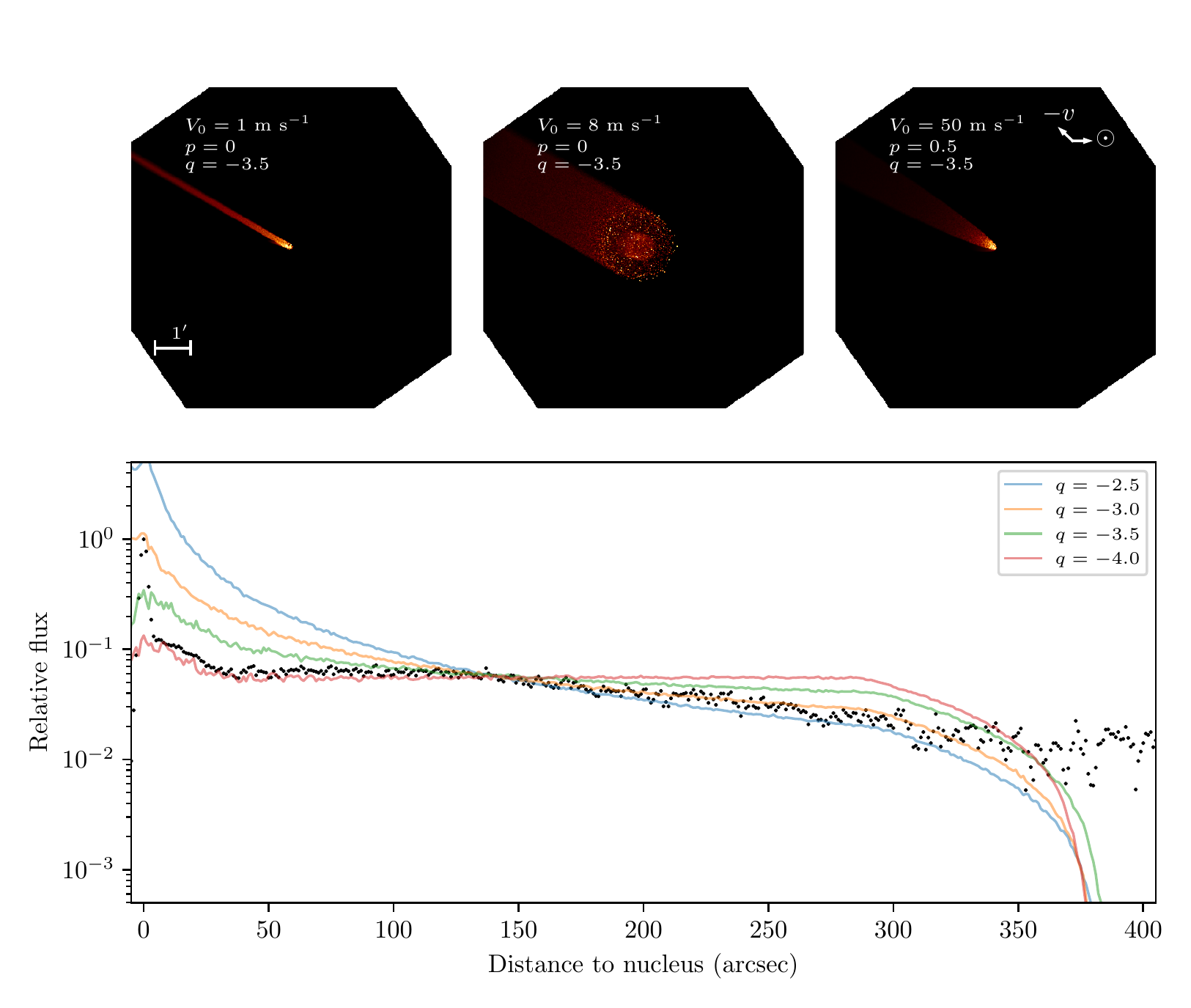}
\caption{Dust tail model for the ZTF image taken on UT 2019 February 8.31. Upper panel: modeled images of three different sets of parameters; lower panel: observed and modeled profile along the tail direction, normalized to the flux level at $140''$ (1.6$\times 10^6$~km at the asteroid) behind the nucleus. Notebook is available \href{https://github.com/Yeqzids/activation_of_6478_gault/blob/master/dust_model.ipynb}{here}. \label{fig:dust-model}}
\end{figure*}

Model images, shown in the upper panel of Figure~\ref{fig:dust-model}, clearly show that only $V_0=1~\mathrm{m~s^{-1}}$ and $p=0$ agrees with the observation. This indicates that the dust grains are released just beyond the gravitational escape speed of Gault. A constant ejection speed through the particle size range is in line with the behavior of other active asteroids with non-sublimation-driven ejection \citep{Jewitt2015}.

Since the tail is straight and narrow, we collapse the observational and model images into 1-dimensional profiles along the tail axis for simple comparison. Note that the tail axis is $31^\circ$ off from the Sun-comet axis in the anticlockwise direction, due to the coupling of gravitational attraction and radiation pressure on the dust grains. To minimize sky noise, the observational profile is derived from summing a $20''$-wide strip along the tail axis, with the width of $20''$ corresponding to the width of the tail.

The profiles, shown in the lower panel of Figure~\ref{fig:dust-model}, reveal a turnover point at $\sim140''$ behind the nucleus, corresponding to particles with $\beta=0.01$ according to the syndyne analysis in Figure~\ref{fig:fp}. Particles with $\beta<0.01$ follow $q=-4.0$ while those with $\beta>0.01$ follow $q=-2.5$ to $-3.0$. A quick check shows that the turnover is also seen in the data taken on 2019 January 28. A change in the power-law of the dust size distribution is interesting but not unusual: similar behaviors have been observed in meteor observations \citep{Ye2014} and in-situ measurements \citep{Gombosi1986, Price2010}, and probably reflect either inherent properties of individual comets or spaceweathering process on different sizes of dust. Note that these materials are ejected during the first event; ejecta from the second event are within $40''$ from the nucleus and do not have as much statistics. We crudely estimate $q=-3.5$ to $-4.0$ for the ejecta from the second event.

\subsection{Mass Loss} \label{sec:an:mass}

The total mass loss can be calculated by: 

\begin{equation}
M_\mathrm{d} = \frac{4}{3} \rho_\mathrm{d} \bar{a} C_\mathrm{e}
\end{equation}

\noindent where $\rho_\mathrm{d}=2900~\mathrm{kg~m^{-3}}$ is the bulk density of dust grains \citep{Carry2012}, $\bar{a}$ is the characteristic grain size for the derived size distribution, which we take $\beta=0.01 \Leftrightarrow 20~\micron$, following the results from \S~\ref{sec:an:dust}, and $C_\mathrm{e}$ is the effective scattering cross-section of the ejecta:

\begin{equation}
\label{eq:cs}
C_\mathrm{e} = \frac{\pi r_\mathrm{h}^2 \varDelta^2}{p_V \Phi(\alpha) a_\mathrm{\oplus}^2} \left( 10^{0.4\Delta m_r} -1 \right) 10^{0.4(m_\mathrm{\odot, r} - m_r)}
\end{equation}

\noindent where $\Delta m_r$ is the brightness excess in $r$, $a_\mathrm{\oplus}=1.5\times10^{11}$~m is the mean heliocentric distance of the Earth, $\Phi{(\alpha)}=0.035\alpha$ is the simple phase function of the target with a phase angle of $\alpha$ \citep[c.f.][]{Li2015}, $m_\mathrm{\odot, r}=-26.9$ is the apparent $r$ magnitude of the Sun \citep{Willmer2018}, $m_r$ is the nuclear brightness of the asteroid, and definitions of all other variables follow the definition in Equation~\ref{eq:v_ej}. By substituting corresponding numbers from the two brightness peaks identified in \S~\ref{sec:obs}, taking $\alpha=21^\circ$ over the observed period and assuming $p_V=0.04$ for the dust \citep{Fulle2010}, we obtain $M_\mathrm{d}=2\times10^7$~kg for the first event, and $M_\mathrm{d}=1\times10^6$~kg for the second event.

\section{Activity Mechanism}

Asteroid activity can be driven by a variety of mechanisms, including sublimation, impact, electrostatics, rotational excitation, thermal stress, or interactions of near-contact binaries \citep{Taylor2014, Jewitt2015}. Our observations rule out sublimation (see \S~\ref{sec:dust-model}) or electrostatics (which operates on $micron$-sized dust). Gault's orbit keeps it relatively far from the Sun, ruling out thermal stress. Three possibilities seem to remain: impacts, rotational excitation, and binary interactions.

Gault is likely between 2--10~km in diameter, as estimated from our $HG$ model. This number is calculated using the canonical relation $D = 1329 p_V^{1/2} 10^{-H/5}$, where $p_V=0.03$--0.6 is the geometric albedo of the Phocaea asteroids \citep{Novakovic2017}. The albedo of Gault is not known. The broadband color derived from pre-outburst ZTF data is slightly more compatible with low-albedo, C-type asteroids than other types of asteroids (see \S~\ref{sec:results}), but this needs spectroscopy to confirm.

Statistical models predict that for asteroids in this size range (i.e. asteroid with diameters of $>2$~km), impact-driven activity disruption occurs more frequently than rotation-driven activity \citep{Marzari2011}. However, the repeated activation of Gault would require multiple impacts, which we consider unlikely\footnote{An impact-driven activity would have been an excellent tribute to Dr. Donald Gault, whom the asteroid was named after; he was a pioneer in the field of impact cratering processes \citep{Schultz1999}.}. If the two events were indeed driven by impacts, a pair of 10--20~m projectiles was needed to reproduce the observed brightness increase and the cross sections of the ejecta \citep[][Figure~16]{Jewitt2015}. A rotation-driven event, though consistent with the observation, occur at a rate of $10^{-6}$ to $10^{-4}~\mathrm{yr^{-1}}$ for the entire population of $>2$~km main-belt asteroids, which is low.

The most likely scenario, a merging near-contact binary, was suggested to drive the activity of  311P/PANSTARRS \citep{Hainaut2014}. A binary system can merge into a contact binary through the loss of angular momentum \citep[known as the binary Yarkovsky-O'Keefe-Radzievskii-Paddack effect, or BYORP, c.f.][]{Cuk2005}, but the merging process can cause fragmentation and dust emissions. Multiple ejection episodes may therefore be caused by different stages in the merging process. The timescale of the BYORP process is $\sim10^5$~yr for near-Earth asteroids \citep{Cuk2007}, or $\sim10^6$~yr for main-belt asteroids after accounting for lower solar radiation at larger distances to the Sun. Assuming $15\%$ binaries among main-belt asteroids \citep[see the discussion in][\S~2.2]{Margot2015}, the rate of binary merges is (number of km-sized main-belt asteroids) $\times$ (fraction of binaries) $\div$ (timescale of BYORP process) $=10^6\times15\%\div10^6~\mathrm{yr}=0.1~\mathrm{yr^{-1}}$, which, for km-sized asteroids, is much higher than the rates of impact- and rotation-driven activities for km-sized asteroids.

A merging binary system is expected to have a characteristic light-curve with a large amplitude \citep{Descamps2008}. The available ZTF data do not allow extracting a rotational light-curve; however, a crude inspection of Figure~\ref{fig:lc} does not show any noticeable scatter caused by large amplitude. This may be explained by the geometry between the orbital plane of the binary and the Earth, as well as the contamination of the ejecta. Future light-curve studies, to be conducted after the dust grains have moved away, should be able to prove or reject this theory.



\section{Summary} \label{sec:sum}

We presented observations of newly-identified active asteroid Gault obtained in the course of the ZTF survey. As of mid-Februrary 2019, Gault exhibited two mass loss events: the first event started on 2018 October 18$\pm5$~days, peaked around 2018 November 7, and subsided around mid-December 2018; the second event started on 2018 December 24$\pm1$~days, peaked around 2019 Jan. 4, and subsided around late-January 2019. The two events released $2\times10^7$~kg and $1\times10^6$~kg of mass, respectively. Gault's color did not change appreciably during the outbursts, with a pre-outburst color of $g_\mathrm{PS1}-r_\mathrm{PS1}=0.50\pm0.04$ and $g_\mathrm{PS1}-r_\mathrm{PS1}=0.46\pm0.04$ during the two outbursts.

We investigated the evolution of dust tail morphology using a dust dynamics model. We found that the ejecta was dominated by dust grains with sizes $\gtrsim10~\micron$, and followed different size distribution: particles at 10--20~$\micron$ follow a power-law of $-2.5$ to $-3.0$, while larger particles following a steeper slope of $-4.0$. These particles were ejected at very low speeds of $\lesssim1~\mathrm{m~s^{-1}}$ regardless of particle sizes, consistent with non-sublimation-driven ejections observed in other active asteroids.

The derived properties were most consistent with an activity driven by either rotational excitation, or merger of near-contact binaries. We showed that from a statistical perspective, a merger of near-contact binaries is more likely to be responsible for the activity of Gault, but additional observations are needed to prove or dispute this hypothesis.

\acknowledgments

We thank the referee for a prompt and helpful review, as well as Man-To Hui for comments. Q.-Z. Ye is supported by the GROWTH project funded by the National Science Foundation under Grant No. 1545949. Bryce Bolin and Lynne Jones acknowledge support from the DIRAC Institute in the Department of Astronomy at the University of Washington. The DIRAC Institute is supported through generous gifts from the Charles and Lisa Simonyi Fund for Arts and Sciences, and the Washington Research Foundation. Funding for the Asteroid Institute program is provided by B612 Foundation, W.K. Bowes Jr. Foundation, P. Rawls Family Fund and two anonymous donors in addition to general support from the B612 Founding Circle (\url{https://b612foundation.org/member_group/founding-circle/}).

This work is based on observations obtained with the Samuel Oschin Telescope 48-inch Telescope at the Palomar Observatory as part of the Zwicky Transient Facility project. Major funding has been provided by the U.S. National Science Foundation under Grant No. AST-1440341 and by the ZTF partner institutions: the California Institute of Technology, the Oskar Klein Centre, the Weizmann Institute of Science, the University of Maryland, the University of Washington, Deutsches Elektronen-Synchrotron, the University of Wisconsin-Milwaukee, and the TANGO Program of the University System of Taiwan.

This work made use of sbpy (\url{http://sbpy.org}), a community-driven Python package for small-body planetary astronomy supported by NASA PDART Grant 80NSSC18K0987.

\facilities{PO:1.2m}
\software{Astropy \citep{Astropy2018}, Jupyter Notebooks \citep{Kluyver2016}, Matplotlib \citep{Hunter2007}, MERCURY6 \citep{Chambers1997}, sbpy, ZChecker \citep{Kelley2019}}

\end{CJK*}
\bibliographystyle{aasjournal}
\bibliography{ms}



\end{document}